\documentclass[12pt, preprint]{aastex}
\usepackage{graphics}

\newcommand{\uo}{\mu_0}

\newcommand{\curl}{\nabla\times}

\newcommand{\diver}{\nabla\cdot}
\newcommand{\grad}{\nabla}

\newcommand{\bb}{\mathbf{b}}

\newcommand{\BB}{\mathbf{B}}
\newcommand{\BO}{\mathbf{B_0}}

\newcommand{\vv}{\mathbf{v}}
\newcommand{\vo}{\mathbf{v_0}}

\newcommand{\kk}{\mathbf{k}}
\newcommand{\xxi}{\mbox{\boldmath$\xi$}}

\newcommand{\wa}{\omega_A}
\newcommand{\wab}{\bar{\omega}_A}
\newcommand{\wm}{\omega_m}
\newcommand{\wt}{\omega^2}
\newcommand{\WB}{\Omega_B}
\newcommand{\ddr}{\frac{d}{dr}}

\newcommand{\Af}{Alfv\'{e}n\ }
\newcommand{\beqn}[1]{\begin{equation} \label{#1}}
\newcommand{\eeqn}{\end{equation}}
\newcommand{\beqna}[1]{\begin{eqnarray} \label{#1}}
\newcommand{\eeqna}{\end{eqnarray}}

\title{Global axisymmetric Magnetorotational Instability with density gradients}
\author{Jesse Pino \and S. M. Mahajan}
\affil{Institute for Fusion Studies, The University of Texas at Austin, Austin,Texas 78712}

\email{pino@mail.utexas.edu}

\begin{abstract}
We examine global incompressible axisymmetric perturbations of a differentially rotating MHD plasma with radial density gradients.   It is shown that the standard magnetorotational instability, (MRI) criterion  drawn from the local dispersion relation is often misleading. If the equilibrium magnetic field is either purely axial or purely toroidal, the problem reduces to finding the global radial eigenvalues of an effective potential. The standard Keplerian profile including the origin is mathematically ill-posed, and thus any solution will depend strongly on the inner boundary.  We find a class of unstable modes localized by the form of the rotation and density profiles, with reduced dependence on boundary conditions.
\end{abstract}

\begin{document}

\section{Background}
\label{Background}

It is often stated that the Magnetorotational Instability (MRI) \citep[][hereafter BH91]{Chandrasekhar:1961,Balbus:1991}
in accretion disks is a `local' instability, i.e. normal modes are driven unstable by the local value of the rotational flow shear.  Implicit in this analysis is the assumption that equilibrium rotation and density vary over a much larger spatial scale than the mode wavelength.  Although it has been shown that short-wavelength linear local MRI modes can drive global turbulence in the nonlinear regime \citep{BH98,2001ApJ...554..534H}, it is worthwhile to study linear instabilities with large radial extent which can contribute more direct angular momentum transfer.  In order to study these modes, it is necessary to use a more complete treatment, including the radial variations of the equilibrium profiles. One arrives at a second-order differential equation, which can be solved for the structure of global eigenmodes.

A major drawback to the study of global modes, aside from the computational complexity, is their strong dependence on boundary conditions, which are often unknown for astrophysical systems.  Previous work  \citep[e.g.][hereafter OP96]{1993A&A...274..667D, OP96}, has included rigid boundary walls to discretize the mode spectrum, but this imposition is arbitrary and unphysical.  Another approach \citep{curry94} is to use pressure constraints to define a boundary, outside of which the field is taken to be vacuum. Appropriate matching conditions are then used. \citet{Kersale03} studied the global MRI problem with inflow, and found that certain boundary conditions gave rise to ``wall modes'' with large growth rates.  In this paper,we show that unstable axisymmetric modes in cylindrical geometry can be described by an effective radial potential. The sign of this potential on the boundary dictates how strongly the mode structure depends on the specific boundary conditions taken. We find that smoothly varying equilibrium rotation and density profiles can localize modes and reduce dependence on the explicit treatment of the boundary conditions.  The dispersion relation for these global modes can differ greatly from that of the local treatment. 

Much previous analytical work on global modes has been carried out using the Boussinseq approximation, which treats both the equilibrium and perturbed density as constant except in the equation of motion. This greatly simplifies the dispersion relation, and the density gradient appears only through the buoyancy term (the Brunt-V\"ais\"al\"a frequency). Although this is an appropriate step in the local analysis \citep{Balbus:1991}, when the effective radial wavelength is of the order of the system size, we must allow for significant density variations over the region in question.  In our analysis the density appears in the mode equation in two additional ways; by allowing the local \Af frequency to change with radius, and by introducing terms proportional to the first and second derivatives of the \Af frequency.  It is well known that density gradients in ideal MHD can lead to both continuous and discrete \Af spectra \citep{Sedlacek:1971}. In the absence of equilibrium flow, these modes lead to damped surface eigenmodes \citep[][]{Chen:1974}. If a free-energy source such as differential rotation is present, they can couple to produce unstable modes. We examine how astrophysically relevant density profiles can serve to stabilize modes as well as help the appropriate imposition of outer boundary condition.   

This paper is organized as follows: In Section \ref{Basic Equations}, we derive the radial global mode equation for incompressible axisymmetric perturbations. By allowing for any rotation and density profile, our analysis remains relevant to the global MRI in accretion disks as well as other rotating systems such as laboratory experiments \citep[e.g.][]{Goodman:2002} and stellar core collapse \citep[][]{2003ApJ...584..954A}. In Section \ref{Global Solutions}, we investigate certain illustrative limits of the system represented by this equation. Finally, numerical results are presented in Section \ref{Numerical Results}, showing how certain rotation profiles can lead to direct localization of these modes.

\section{Basic Equations}
\label{Basic Equations}
The equation of motion for an MHD plasma is:
\beqn{eom}
\rho \left(\frac{\partial \vv}{\partial t} + (\vv\cdot \grad)\vv\right) = -\grad P + \frac{1}{\mu_0}(\curl\BB)\times\BB-\grad \Phi_g
\eeqn

Where $\Phi_g$ is the gravitational potential and $P$ is the scalar pressure. The magnetic field evolves according to Maxwell's Equation:
\beqn{max}
\frac{\partial \BB}{\partial t} = \curl(\vv\times\BB)
\eeqn
These equations, along with the divergence condition on the magnetic field, admit a rotating cylindrical equilibrium of the form $\BO = (0,r \WB(r) ,B_z(r) )$ and $\vo = (0,r \Omega(r),0)$. In this equilibrium, we can write (\ref{eom}) as the Euler equation:
\beqn{euler}
\rho \left(r \Omega^2 \hat{\mathbf{r}}-\grad \Phi_g\right) - \grad P_0+ \frac{1}{\uo}(\curl\BO)\times\BO = 0 
\eeqn
The specification of a density, magnetic field, and rotation profiles will determine the pressure up to a constant. We can then define a (local) adiabatic exponent through the identification $P=\kappa \rho^{\gamma}$. 

 If none of the equilibrium quantities depend on the height $z$ or angle $\theta$, we can take Fourier transforms in the axial and azimuthal directions. Neglecting the perturbed gravitational potential $\delta \Phi_g$, the equations for the normal modes of Lagrangian perturbations, ($\xxi=\xxi(r)e^{ i(k_z z+m \theta -\omega t)} $) to this equilibrium are \citep{RevModPhys.32.898,1979MNRAS.187..769C}:
\beqn{L-1}
-\omega^2 \rho \xxi + 2 i \rho (\vo \cdot \grad)\xxi - \mathcal{F}(\xxi) = 0 \ \ ,  
\eeqn
Where
\beqna{L-2}
\mathcal{F}(\xxi) &=&    \grad(\gamma \rho \diver \xxi + (\xxi\cdot\grad)P) + \diver(\rho \xxi)\grad\Phi_g \nonumber \\  
 &\ +& \grad(\BO\cdot \bb)+(\BO \cdot \grad)\bb +(\bb \cdot \grad)\BO  \\
 &\ + & \diver(\rho \xxi (\vo \cdot \grad)\vo - \rho \vo  (\vo \cdot \grad)\xxi) \nonumber.
\eeqna
The perturbation of the magnetic field is 
$\bb = \curl(\xxi \times \BO) $. 
 For incompressible  ($\diver\xxi = 0$) perturbations, eq. (\theequation) can be written as two scalar equations:
\beqna{mne0-1}
(\rho \wm^2 - \wab^2) \ddr \psi_T &=& \left[ (\rho \wm^2 - \wab^2)(\rho \wm^2 - \wab^2 - 2 \rho r\Omega \Omega'+ 2 r \WB \WB' + \rho N^2) \right. \\ 
&-& \left.4(\rho \Omega \wm +\WB \wab)^2\right]\xi_r  
 + 2 \frac{m}{r} (\rho \wm \Omega +\WB \wab)\psi_T \nonumber
\eeqna
\beqn{mne0-2}
(\rho \wm^2 - \wab^2)\frac{1}{r} \ddr(r \xi_r) = -  2 \frac{m}{r} (\rho \wm \Omega +\WB \wab) \xi_r + \left( \frac{m^2}{r^2}+k_z^2 \right)\psi_T
\eeqn

Where $\wm = \omega-m \Omega, \wab = k_z B_{0z}+m\WB$,  $\psi_T$ is the total perturbed pressure (gas plus magnetic), and  $N^2 =- \frac{\rho'}{\rho}(r \Omega^2- \grad \Phi_g)$ is the Brunt-V\"ais\"{a}l\"{a} frequency. In the present treatment, we allow $\rho$ to vary significantly over the region under consideration; as long as we restrict our analysis to incompressible perturbations, the above equations are still valid. Through density variation, the local \Af  frequency can change even when the equilibrium magnetic field is constant. The density length scale $L_{\rho} = ( d (ln \rho)/dr)^{-1}$ is taken to be much larger than the ion Larmor radius (drift waves are not considered).

\subsection{Axisymmetric modes}
\label{Axi}
For this paper, we restrict our consideration to axisymmetric modes (m=0). Then Eqs.~(\ref{mne0-1}) and (\ref{mne0-2}) can be reduced to a single second-order differential equation in the radial coordinate:
\beqn{m0eq}
\ddr \left[F(r) \frac{1}{r} \ddr (r \xi_r) \right] - k_z^2 \left[ F(r)- 2\rho r \Omega \Omega' +2 r \WB\WB'+ \rho N^2-\frac{4 (\rho \Omega \omega+\WB \wab)^2}{F(r)}\right]\xi_r =0
\eeqn

Where $F(r)=\rho(r) \omega^2 - \wab^2$, and $\wab^2= k_z^2 B_{z0}^2/\uo$. This equation describes the standard MRI in the limit $\rho' \rightarrow 0$, and the gravitational Interchange Instability in the limit $\Omega \approx 0$, $\Omega' \ne 0$.  If there is no equilibrium rotation in the system, the mode equation is the cylindrical form of the well known differential equation for surface Alfv\'{e}n waves \citep{Sedlacek:1971}. There is a continuum of stable oscillations at each frequency $\omega = \wa(r) = \kk\cdot\BO/\sqrt{\uo\rho(r)}$; each frequency is strongly localized around the characteristic radius where $F(r)$ vanishes.  These modes overlap spatially and give damping proportional to 1/t.  In addition, there exists a discrete spectrum of surface modes with position-indepenent frequency. This is the phenomenon of damped resonant absorption, with weak damping for sharp variations in density \citep{HasegawaUberoi}.  Finite Larmor radius terms couple these modes to the Kinetic \Af Wave (KAW) \citep[][]{Mahajan:1984}, which will not be addressed in this paper.   

The mode frequency $\omega$ enters the the differential equation (\ref{m0eq}) only through $F$ and the last term in square brackets. If either $B_{z0}$ or $\WB$ vanishes, only $\wt$ appears. Since all other terms are real, the eigenvalues $\omega$ must then be purely real or imaginary \citep[][]{Chandrasekhar:1960}.  Although the presence of velocity shear makes the evolution equation non-Hermitian, when restricted to the normal mode solutions of purely axial or toroidal fields, we obtain a fully Hermitian eigenvalue problem (provided that the equation is well-behaved at the boundaries). This allows for a significant simplification in the search for global modes. Before proceeding, we examine the local limit of the above mode equation. 

\subsection{Local Dispersion Relation}
In order for modes to be spatially oscillatory, we must have $\xi''/\xi < 0$. If the radial variation of the equilibrium quantities is small with respect to the scale of the perturbation, we can take an expansion $\xi(r)= 1 + \beta_r(r-r_0)^2/2$ in equation (\ref{m0eq}). Solving for $\beta(r=r_0)$, and finding which values of $r_0$ make $\beta_r < 0$, we obtain:
\beqn{betaeq}
\beta_r =k_z^2\left(1 - \frac{2\rho r \Omega \Omega' -2\rho r \WB \WB' }{F} + \frac{\rho N^2}{F} -\frac{4 (\rho \Omega \omega + \WB \wab)^2}{F^2}\right)+\frac{1}{r^2}-\frac{1}{r}\frac{F'}{F},
\eeqn
where here all quantities are taken at their local values.
In the Boussinesq limit ($F' = 0,\ N^2\neq 0$), this reproduces the local MRI dispersion relation of BH91, if we identify $\beta_r \rightarrow - k_r^2$ and take $ k_r,k_z \gg 1/r$. Thus the global analysis and the local analysis agree in the appropriate limit. However, this equation provides no indication as to which radius should be used when applying this criterion, or what to do if it is satisfied in some places and not in others. The local instability criterion can be useful in locating the region containing the most unstable mode, which for the MRI generally occurs near the point of greatest shear. If the mode has radial extent comparable to the equilibrium variation, the full global analysis can lead to results quite different from this local criterion.

\subsection{Effective Potential}
\label{Effective Potential}
When the equilibrium magnetic field is purely axial, eq. (\ref{m0eq}) only admits modes with real $\wt$. If the \Af term $F(r) = \rho \wt-\wab^2$ has the same sign for all $r$ in the domain, we can make the substitution $y= \sqrt{\pm r F} \xi_r$. For purely growing modes, $F(r)$ is negative for all $r$, regardless of the form of the density profile. We arrive at: 
\[
\frac{d^2 y}{dr^2} - V(r,\omega)y =0 
\]
\beqn{effpot}
V(r,\omega) = k^2\left(1 - \frac{2\rho r \Omega \Omega' }{F} + \frac{\rho N^2}{F} -\frac{4 \rho^2 \Omega^2 \omega^2}{F^2}\right) +\frac{3}{4r^2} + Q(r)\eeqn
\[Q(r)= - \frac{1}{2r}\frac{F'}{F} + \frac{1}{2}\frac{F''}{F}- \frac{1}{4}\frac{F'^2}{F^2}\]

The problem becomes one of finding the zero energy solutions of the frequency dependent  ``effective potential'' $V(r,\omega)$ \citep{IFSR-1133}. If the potential is positive everywhere, the solutions are monotonic, and it is impossible to construct a global solution satisfying both boundary conditions. It is therefore necessary that $V(r,\omega) <0$ in some region in r for a global mode to be possible. There are two distinct ways for this to occur: 
\begin{enumerate} 
\item[1)] $V(r,\omega)$ is negative all the way up to one or both of the boundaries of the region under consideration. This gives rise to boundary-localized `wall'  modes like the ones found in OP96. Any change of the boundary condition will drastically affect the mode structure and frequency spectrum. The Keplerian flow profile is always of this type for the inner boundary, as we shall see below. 
  
\item[2)] $V(r,\omega)$ has a minimum which is less than zero at some radius, but is positive elsewhere. In this case, the region of oscillation is localized by the potential well, and the mode is spatially evanescent outside the well. The boundaries play a reduced role in the mode structure, although they can still be important. The local stability criterion may be satisfied over significant portions of the disc yet unstable modes can exist which are localized by the effective potential well. The most unstable modes are the ones with no zero crossings; these modes tend to have a greater radial extent and thus a greater chance of carrying radial angular momentum. 
\end{enumerate}

If the equilibrium magnetic field is purely toroidal, the effective potential becomes: 
\beqn{effpottor}
V(r,\omega) = k^2\left(1 - \frac{1}{\wt}(4  \Omega^2+2 r \Omega \Omega' -N^2 - 2 r \WB \WB' /\rho ) \right) +\frac{3}{4r^2} + Q(r). 
\eeqn
Since the equilibrium magnetic field is perpendicular to $\kk$, the \Af term $F\to\rho \omega^2$, and the only coupling to the magnetic field is through the equilibrium magnetic shear $2 r \WB \WB'$. If the modified Rayleigh criterion
\[4  \Omega^2 + 2 r \Omega \Omega' -N^2 -2 r \WB \WB' /\rho> 0,\]
is satisfied, the potential is always positive for purely growing modes, and the system is stable to $m=0$ perturbations (i.e. there are no global axisymmetric MRI modes). In the absence of rotation, this is the Tayler ``pinch'' stability criterion \citep[][]{1973MNRAS.161..365T}. The current-free configuration $\WB = \beta r^{-2}$ is always stabilizing.  In what follows, we examine only purely axial magnetic fields, and defer consideration of toroidal fields to a later paper examining non-axisymmetric disturbances. 

%--------------
\section{Global Solutions}
\label{Global Solutions}
We begin by investigating various limits of the global mode equations analytically. 
\subsection{Rigid Rotation}
\label{rigid}
If  $\Omega=\Omega_0,\, \BB=B_{z0}$, and the density is constant, the mode equation reduces to ($F_0 = \wt - \wab^2/\rho_0=const.$):
\beqn{rigid-1}
{d\over dr}{1\over r}{d\over dr}r\xi_r -k_z^2\xi_r =-\frac{4k^2 \omega^2 \Omega_0^2}{F_0^2}\xi_r ,
\eeqn
allowing a family of solutions
\[ \xi_r =A J_1(\mu r) + B Y_1(\mu r), \]
describing shear Alfv\'en waves in a rigidly rotating homogeneous plasma \citep[][]{HasegawaUberoi}. Here $\mu$ is to be interpreted as an effective radial wavenumber obeying:
\beqn{rigid-2}
\mu^2=k_z^2\left(\frac{4 \wt \Omega_0^2}{F_0^2}-1\right) \, .
\eeqn
The values of $\mu^2$ are determined by matching the solutions of Eq .~\ref{rigid-1} to the imposed boundaries. This results in a (boundary-dependant) discrete spectrum of stable eigenmodes when $\mu^2>0$ \citep[][]{1993A&A...274..667D}.

If the dispersion relation of equation (\ref{rigid-2}) gives a negative value for $\mu^2$, the solution is a linear combination of the modified Bessel functions $I_1(|\mu| r)$ and $K_1(|\mu| r)$. When the rotation frequency is constant throughout the entire domain, there can be no global mode satisfying both boundaries, as both solutions to eq. (\ref{rigid-1}) are monotonic. If only a portion of the domain is subject to rigid rotation (the effective potential is positive in that region but negative elsewhere), the modified Bessel functions provide suitable limiting forms.  In particular,  when either the density or the rotation are small for large r, we obtain $|\mu| \approx k_z$ (vacuum solution).  We will use this result in Section \ref{Numerical Results} to provide interior and exterior boundary matching conditions for modes localized by the form of the equilibrium profiles.%-----------------------
\subsection{An Exactly Solvable Profile}
\label{aor}
For the next limit, we investigate a system with differential rotation for which we can find exact solutions. Take constant density $\rho=\rho_0$, let
\beqn{aor1}
\Omega^2=\Omega^2_0\left[\frac{\alpha}{r}+\beta\right],
\eeqn
and take the magnetic field to be uniform in the $\hat{z}$ direction. For this profile $\Omega'< 0$ if $\alpha> 0$. We have deliberately chosen this form 
so that the result for rigid rotation can be obtained by letting $\alpha\to 0, \beta\to 1$.  A pure power law is obtained when $\beta \to 0$. As we have stressed above, any rotation profile may be obtained by specifying an appropriate equilibrium pressure. The effective potential equation
\beqn{aor2}
 \frac{d^2 y}{d r^2} -\left[\frac{3}{4 r^2}+k_z^2\left(1-\frac{4\beta \Omega^2_0\omega^2 } {F_0^2}\right)+
k_z^2 \frac{\alpha\Omega^2_0}{F_0 r}\left(1-\frac{4\omega^2}{F_0}\right) \right]y =0,
\eeqn
may be written in the standard Whittaker form \citep[][]{AMS55}
\beqn{aor3}
\frac{d^2 y}{dr^2} -\left[p_0-\frac{q_0}{r}+\frac{3}{4r^2}\right]y =0,
\eeqn
where 
\[p_0 = k_z^2\left(1-\frac{4\beta \Omega^2_0\omega^2} {F_0^2}\right) \, \, , \, \,
 q_0 = -\frac{\alpha k_z^2 \Omega^2_0}{F_0}\left(1-\frac{4\omega^2}{F_0}\right).\]
On a semi-infinite domain $r\in(0,\infty)$, it allows there are well-behaved solutions ($\Phi$ is the Kummer Function):
\beqn{aor4}
y_n=\xi_n \sqrt{-F_0 r}=A\ r^{3/2}  e^{-\sqrt{p_0}\, r} {\Phi} \left[{3\over 2}-
{q_0\over 2\sqrt{p_0}},\ 3,\ 2\sqrt{p_0}\, r\right].
\eeqn
This solution requires $p_0> 0$, which is satisfied for purely growing modes (all modes if $\beta\to 0$). The eigenvalue condition arises from the need for the displacement and its radial derivative to be bounded at both $r\to 0$ and $r \to \infty$; the latter demands that the Kummer series terminates. This happens when  
\beqn{aor5}
{3\over 2}-{q_0\over 2\sqrt{p_0}}=-n,
\eeqn
yielding the dispersion relation,
\beqn{aor6}
\frac{\alpha k_z^2 \Omega^2_0}{F_0}\left(1-\frac{4\omega^2}{F_0}\right)=
-k_z\sqrt{1-\frac{4\beta \Omega^2_0\omega^2}{F_0^2}}\ (2n+3).
\eeqn

If $\beta\to 0$, we can solve equation (\ref{aor6}) for the frequency:
\beqn{aor7}
\wt_n = \wa^2 + \frac{k_z \alpha \Omega_0^2}{2(2n+3)}\left(3 \pm  \sqrt{9  +16\frac{\wa^2 (2n+3)}{k_z \alpha \Omega_0^2}}\right). \eeqn
 From eq. (\ref{aor7}), we derive the instability criterion:
\beqn{aor8}
\alpha\Omega_0^2>k_z v_a^2\ (2n+3)
\eeqn
where $v_a=\wa/k_z$.  

\begin{figure}[ht]
\plotone{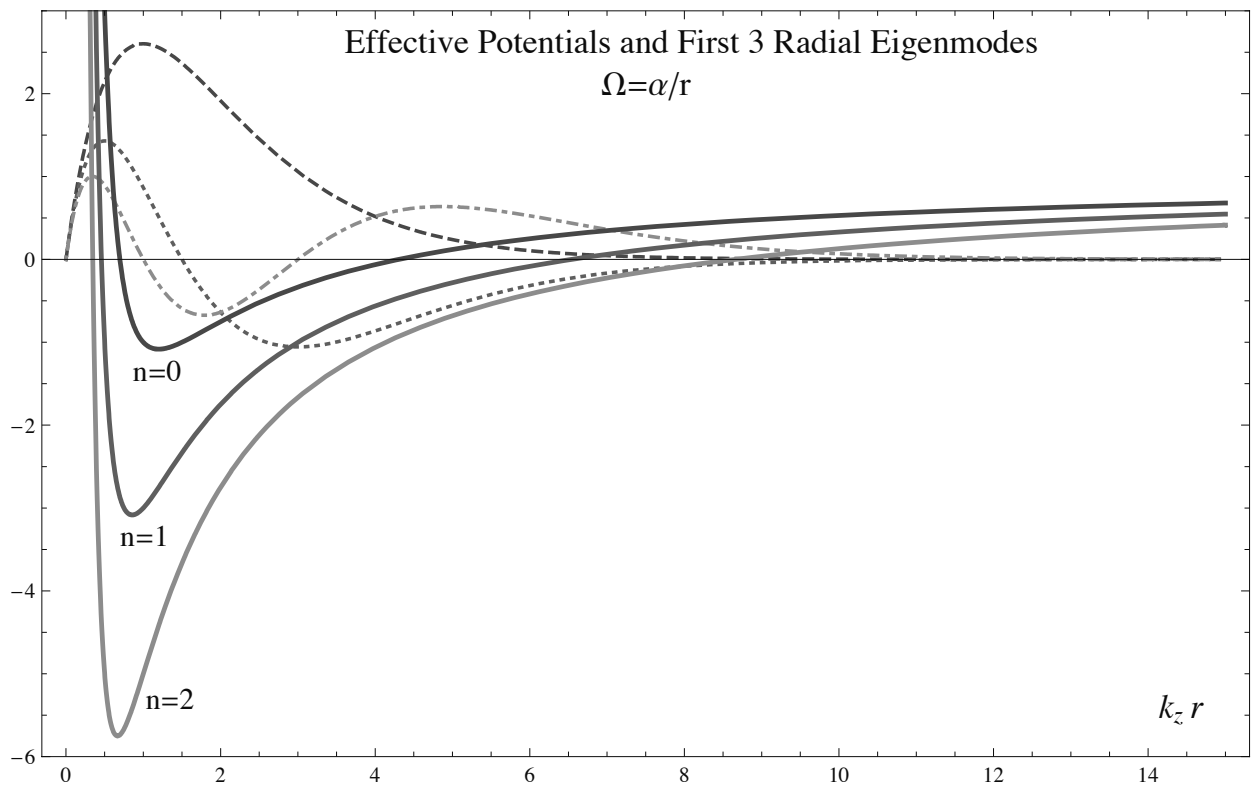}
\caption{ The profile $\Omega^2 = \alpha/r, \rho=const.$, considered in section \ref{aor}, admits a discrete spectrum of eigenmodes. For each radial quantum number $n$, there exists an effective potential (solid lines), which supports a global mode satisfying the evanescent boundary conditions (dashed lines).  As $n$ increases, the potential well gets deeper, and the mode extends over a greater area. For the unstable branch, higher n corresponds to smaller growth rate. }
\label{alphaoverr}
\end{figure}

\begin{figure}[ht]
\plotone{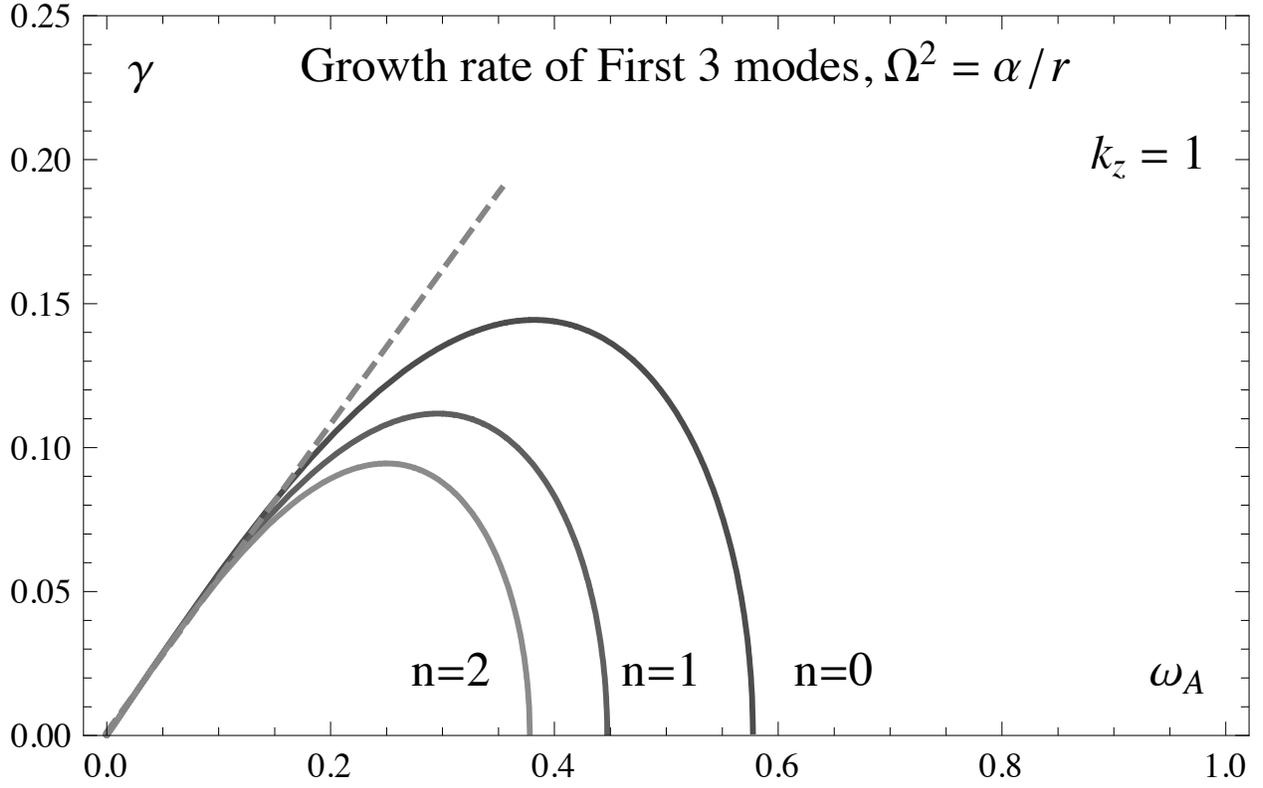}
\caption{ Growth rate vs. \Af frequency for the first three radial eigenmodes of the profile $\Omega^2 = \frac{\alpha}{r}, \rho=\rho_0, k_z=1.$ The lowest order radial mode is the most unstable for all magnetic field values. As the magnetic field strength is increased, the growth rate of each mode first increases and then decreases until the mode is stabilized. }
\label{alphadisp}
\end{figure}

The radial quantum number $n$ appears explicitly in the spectral relation. The first three modes and associated effective potentials are plotted in Figure~\ref{alphaoverr}.
For small $k_z$, equation (\ref{aor8}) provides a more severe constraint on $\Omega'$ (measured by $\alpha$) than the local criterion $k_z^2 v_a^2 < 2 r \Omega\Omega'$ (BH91). 
We see that as the radial mode number $n$ rises, the instability criterion becomes harder to satisfy, and the higher order radial modes are less unstable. Thus the converse of eq. (\ref{aor8}) with $n=0$ can be taken as a necessary condition for global stability. 
 In Figure \ref{alphadisp}, we plot the growth rate versus \Af frequency for the first three radial modes, for $k_z=1$. We see that when $\wa$ is small, we have $\gamma \sim \wa/\sqrt{3}$ for all modes. As the magnetic field is increased, the lowest order modes remain the most unstable. The growth rate for each mode reaches a peak value of $\gamma=\sqrt{\alpha}\Omega_0 /(4\sqrt{2n+3})$ at the \Af frequency $\wa = 7\sqrt{\alpha}\Omega_0 k_z /(4\sqrt{2n+3})$.  The growth rate then decreases as $\wa$ increases. There exists a critical magnetic field strength above which each mode ceases to be unstable, with the lowest order radial mode persisting to the highest field value. 

\clearpage

\clearpage

In this case, the instability criterion, though both qualitatively and quantitatively different from the ``local" criterion, can indeed be satisfied 
for $\Omega'<0$. The eigenvalue problem is well posed;  well-defined square integrable  
eigenfunctions are associated with unstable modes. The discretization is entirely defined by satisfaction of the outer boundary condition. If we had instead imposed hard boundaries at some radii  $r_1$ and $r_2$, we would arrive at a different spectrum by including the second solution to Whittaker's equation, which blows up near the origin. This illustrative case shows the perils of a rotation profile which is unbounded near the origin (making the inner boundary very important), as well as one which has constant density and slow fall-off of rotation out to large radius (making the outer boundary important). 
%----------------------
\subsection{Keplerian Profile}
When the equilibrium pressure and self-gravitation of the plasma are negligible, we obtain the Keplerian case, $\Omega = \Omega_0 r^{-3/2}$ ($\Omega_0^2=GM/R_0^3$), whence the Brunt-V\"ais\"al\"a frequency goes to zero.  This is the case most often used in the study of thin accretion disks \citep[][]{BH98,FKR:2002}. The effective potential is 
\[V_{Kep} (r,\omega) = V(r,\omega) = k^2\ - \frac{ \lambda(r,\omega)}{r^3} +\frac{3}{4r^2}+Q(r,\omega) ,\]
where we have defined 
\[\lambda(r,\omega) = k^2 \Omega_0^2 \frac{ \rho^2 \omega^2+ 3\rho \wab^2}{F^2},\] 
and $Q(r)$ is defined as in the previous section. If the density is constant, $Q(r)=0$, and $\lambda$must then be a positive constant if there is to be a spatially oscillatory MRI mode. The effective potential  $V_{Kep}$ then has no potential well-- it takes its minimum negative value on the inner boundary (Case 1 described in Section \ref{Effective Potential}). If the region contains the origin, the point $r=0$ is an essential singularity ($V_{Kep}\sim\lambda/r^3 \to -\infty$). As the The solutions to the differential equation near this singularity, although bounded, have divergent first derivatives.  The problem of Keplerian rotation including the central point is mathematically ill-defined in the constant density case, and no global MRI mode can be supported.   

We now consider the effect of non-constant density on purely growing global modes ($ \gamma^2=-\wt>0$). We assume that for astrophysically relevant cases, $\rho$ is bounded and that $\rho'(r) < 0$ (the convective stability criterion is satisfied). The \Af frequency becomes a function of position, and $\lambda$ can become negative if  $\gamma^2 > 3 \wab^2/\rho$ (recall that $\wab=k_z B_{0z}/\sqrt{\uo \rho_0}$). This leads to the conclusion that the maximum growth rate of the local MRI in a Keplerian flow profile is $\gamma_{max} = \sqrt{3} \wa(r)$. In the global case, however, a mode with a given growth rate may be spatially evanescent in one region but osciliatory in the other-- a potential well is created. If $\lambda$ remains negative as $r\to 0$, the modes will be well-defined. Since this criterion depends on the value of $\wt$, there will always be some growth rates for which $\lambda >0$ sufficiently close to the origin. The complete spectrum will still be ill-defined.

Density variation also introduces terms to the effective potential (eq. [\ref{effpot}]) which depend on the gradient of the \Af term $F(r)=\rho \wt-\wab^2$. If the density has a power-law profile in some region ($\rho = \rho_0 r^{-a}, a>0$), we find that the part of $V$ that is entirely due to density variation is
\[Q(r) = \frac{a^2}{2 r^2 (1+W r^a)^2} \left(\frac{1}{2}+W r^a\right)>0, \] 
where $W =  \wab^2/(\rho_0 \gamma^2)$.  The contribution is therefore positive for unstable modes. For small r, we have $Q \sim 1/r^2$. This inhomogeneity induced term does not diverge as fast as the Keplerian term, and has little effect on the inner boundary. An exponential density drop $\rho \sim e^{-r/\lambda}$ gives similar results. We conclude that density variation cannot remove the essential singularity that arises due to Keplerian rotation.

The density gradient term $Q(r)$ can be negative if the density profile is locally linear and drops near to zero quickly. If $\rho=\rho_0(1-\Delta(r-r_0)/a)$ between $r_0$ and some $r_2=r_0+a $, then
\[ Q(r) = \frac{\Delta}{2 a}\frac{1}{1+W - \frac{\Delta}{a}(r-r_0) }\left(\frac{1}{r} - \frac{\Delta}{2 a}\frac{1}{1+W  -\frac{\Delta}{a}(r-r_0) }\right),\] which is negative for 
\[(r-r_0) > \frac{2 a }{3 \Delta} (1+W) - \frac{r_0 }{3}.\]
This has an interesting consequence for models which use the constant density approximation over much of the range and then assume zero density outside some boundary \citep[e.g.][]{curry94}. While discontinuity matching may be used, a sharp density drop over a small region can lead to very unstable wall modes, since smaller values of W are more likely to have negative $Q(r)$ over a wider range. Also, the further out the density drop-off occurs, the more negative this term will be (if $\Delta$ is small and $r_0$ is large). If the rotation is sub-Keplerian, a negative radial density gradient implies $N^2 < 0$, so $N^2/F >0$ for unstable modes, and the buoyancy terms can help to mediate this effect. 

We see that density gradients in most cases serve to `shut off' modes that exist for constant density. Both the local \Af variation and the positive $Q(r)$ terms serve to shrink the region over which the effective potential remains negative. Eventually for a given $k$ and $ \wab$, the density profile becomes so steep that the most unstable mode is no longer supported. Since lowering the density has the effect of raising the local \Af frequency, we are in effect raising the `average' \Af frequency for the mode. This phenomenon mimics an effective raising of the magnetic field, which is known to shut off MRI modes. 

\section{Numerical Results}
\label{Numerical Results}
In this section, we numerically examine how modified rotation and density profiles can localize and discretize the unstable radial modes. For given $\Omega(r)$ and $\rho(r)$, we select the axial wavenumber $k_z$ and fiducial \Af frequency $\wab$, and use a shooting and matching code to find the growth rates and radial structures of the unstable modes. At the inner boundary, starting conditions for the shooting routine are obtained by assuming that there is some radius below which both the density and the rotation can be taken constant, ($\rho=\rho_0$ and $\Omega = \Omega_0$). The bounded solution in that region is then the modified Bessel function $I_1(|\mu_0| r)$, as discussed in Section \ref{rigid}. The outer boundary is handled in a similar fashion. Care is taken to choose the outer boundary far enough out such that the resulting growth rate and mode structure so obtained by does not change appreciably when the boundary is moved.

\subsection{Modified Keplerian Profile}
We do not attempt to model real accretion disc boundary layers near the inner object; to do so would require a full treatment of pressure, accretion inflow, etc. \citep[see, e.g.,][]{1993Ap&SS.204....9R}. Rather, we wish to examine the simplest rotation profiles relevant to global MRI. Since the MRI is driven by strong shear, it makes sense to examine profiles that are flat for small radii, and which smoothly transition to rapid fall-off for large radii.  To this end, we take a general form 
\beqn{rotprof}
 \Omega(r) = \frac{\Omega_0}{1+((r-r_1)/R_0)^{s}},\ \ \  r>r_1 \ \ ; \ \ \Omega=\Omega_0, \ \  \ \ r<r_1 
\eeqn
For the case $s=3/2, \ r_1=0$, this profile approaches Keplerian for $r\gg R_0$, but tends to a constant $\Omega_0$ at the origin. This corresponds to a simple model of an accretion disk where the pressure support is only significant below some radius $R_0$, the inner motion corresponding to that of a rigid rotor. The shear is maximized near $r\simeq 0.342 R_0$, as opposed to the true Keplerian case where the shear remains unbounded as the origin is approached. The density is taken to be constant (we shall examine the effects of varying density below). We normalize frequencies to the central rotation frequency $\Omega_0$, and lengths to the fiducial radius $R_0$. The density and magnetic field are normalized such that $v_A R_0^{-1}= \Omega_0$ for $v_A$(normalized)=1. 

\begin{figure}
\plotone{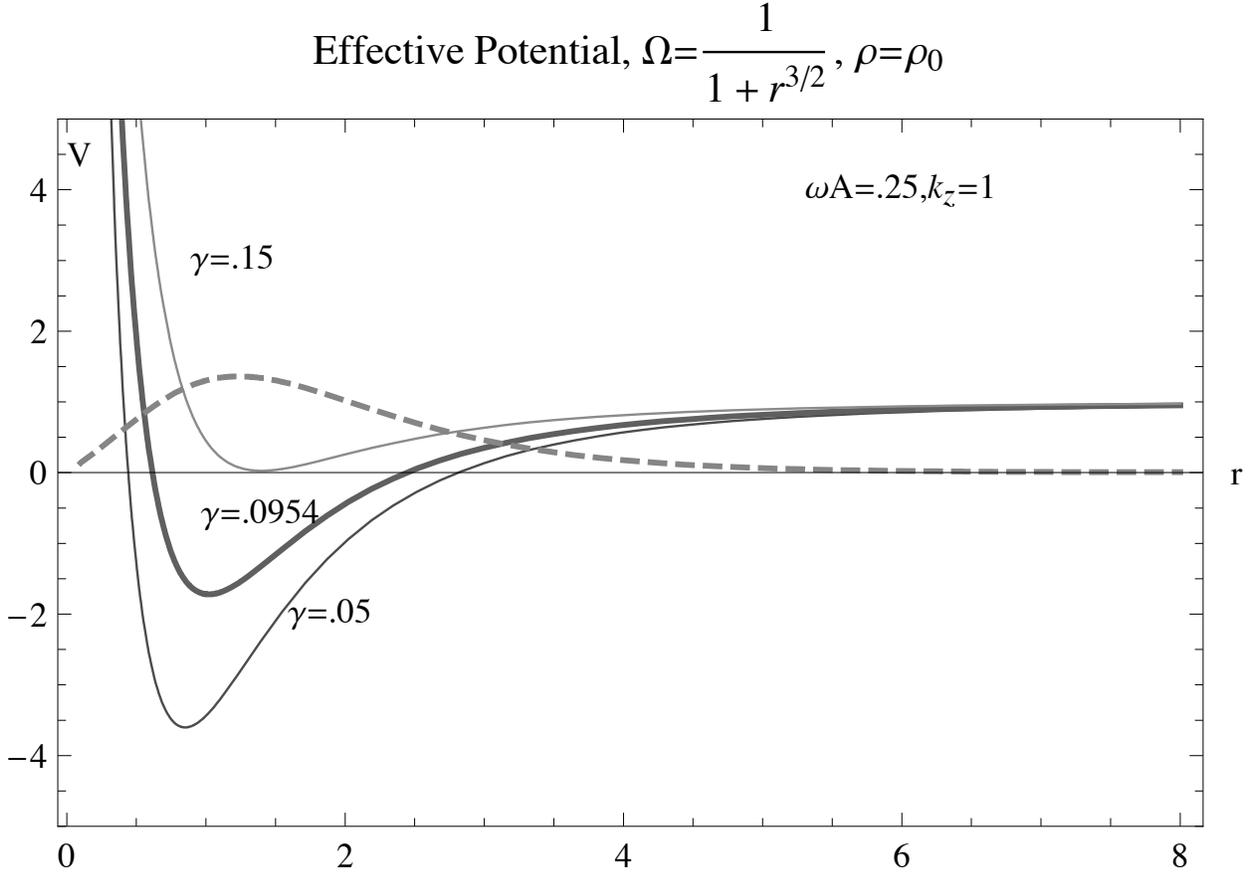}
\caption{ Effective Potential for the rotation profile $\Omega = \Omega_0/(1+r^{3/2}),\ \rho=\rho_0.$ For $k_z=1,\ v_A = .25$, we plot $V(r,\omega)$ for three different values of $\gamma\ (=-i \omega)$. As $\gamma$ increases, the potential becomes less negative. When  $\gamma=0.0954$, the potential supports the lowest order (n=0) discrete radial eigenmode (dashed line).  Even though there is a negative potential for larger growth rates, no eigenmode exists which satisfies both boundary conditions; $\gamma=0.0954$ is the most unstable mode for these $k_z$ and $v_A$ values and all higher $n$ modes have smaller growth rates.. For $\gamma$ above $\sim 0.15$, the potential is no longer negative anywhere.}
\label{effpotkep1}
\end{figure}

The effective potential for this profile is shown in Figure \ref{effpotkep1} for three values of the growth rate $\gamma$, and for $k_z = 1,\ v_A = .25$. Modes with more rapid radial oscillation have smaller growth rates-- this corresponds to a deeper potential well. When the well is very deep, the radial wavelength is small, and a local treatment becomes valid, but only near the bottom of the well. The most unstable mode has no nodes ($n=0$).  Although the local criterion for instability can be satisfied at some radii for larger growth rates, no eigenmodes exist with these larger growth rates which satisfy the evanescent boundary conditions.   

\clearpage
\clearpage

\begin{figure}
\plottwo{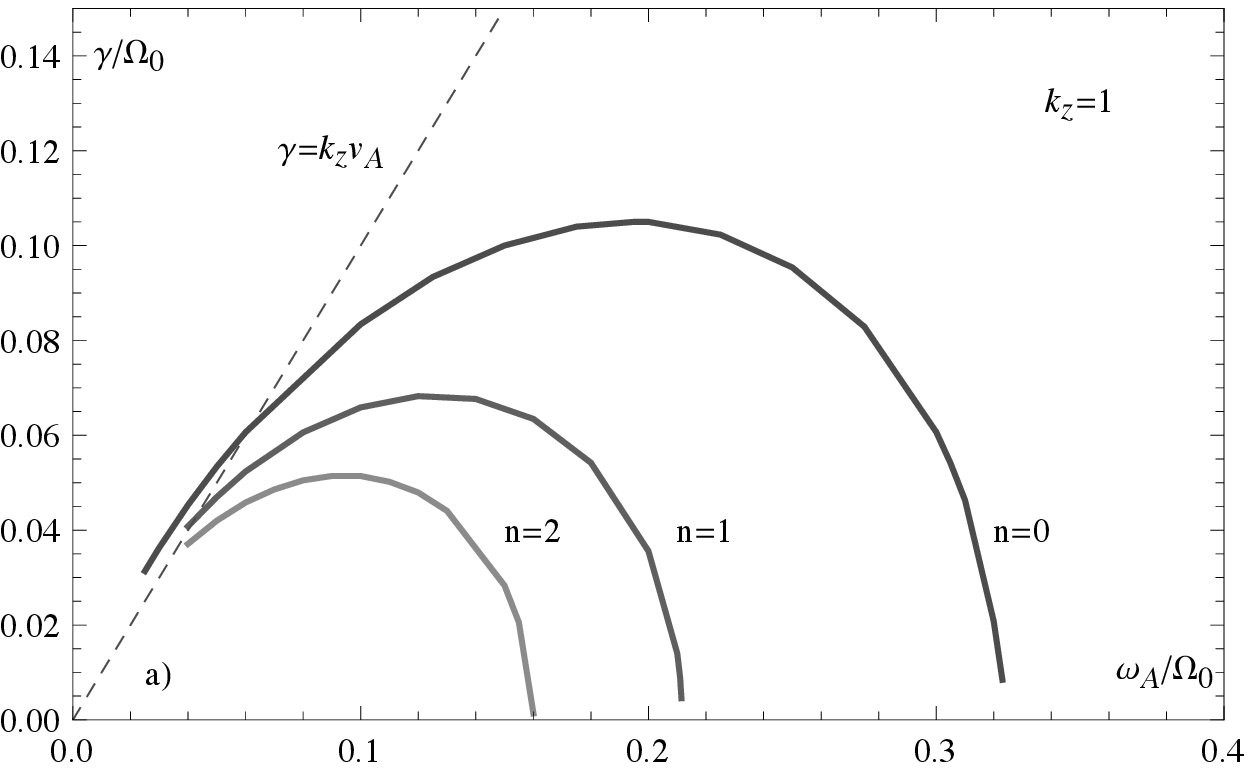}{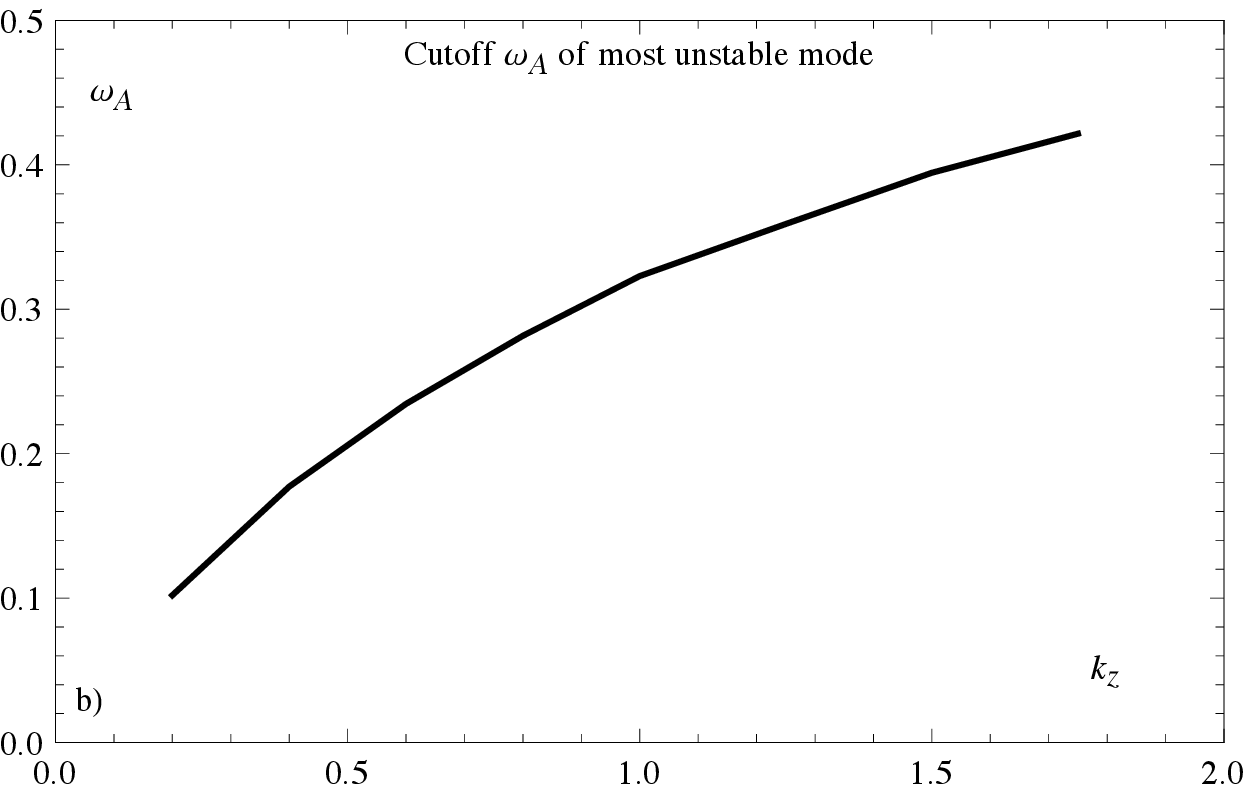}
\caption{ a) The growth rate of the first three radial eigenmodes versus magnetic field strength for  the rotation profile $\Omega = \Omega_0/(1+r^{3/2})$ at fixed $k_z=1$. We see that as the local \Af frequency increases, we transition from a shear like mode $\gamma \sim k_z v_A$ to a cutoff. The maximum growth rate for this wavenumber is $\gamma_{max}=.105$, occurring at $k_z v_A = .2 \Omega_0 $. b) The critical \Af frequency (in units of the central rotation frequency) above which no unstable global modes are supported as a function of axial wavenumber. Since the slope of this plot is less than linear, the critical field strength $\bar{v}_{A crit}$ decreases with increasing $k_z$.  }
\label{cutoff1}
\end{figure}

The dependence of the growth rate on the strength of the background magnetic field is qualitatively similar to the case considered in Section \ref{aor}, and is plotted in Figure \ref{cutoff1}. For small values of the magnetic field, we find that the unstable radial modes are very close together in growth rate and lie close to the shear \Af wave dispersion relation $\gamma=k_z v_{Az}$. For the parameters given, the most unstable mode reaches a maximum at $\wa=0.2,\ \gamma_{n=0}=.105$. At this field strength, the $n=1$ mode has $\gamma_{n=1}=.036$, implying that after 1 rotation, the $n=0$ mode dominates by a factor of $\sim e^{3}\approx 20$. For stronger background fields, the growth rates diminish, and the higher radial order modes are stabilized. Eventually, the magnetic field becomes so strong that even the lowest order radial mode is no longer supported-- the effective potential is not sufficiently deep to support a radial eigenmode. Also plotted is the numerically determined critical stability boundary $\omega_{A\ crit}$ for the most unstable mode as a function of the vertical wavenumber (in the regime $k_z\sim R_0^{-1}$). Note that in this range, the mode can have a significant radial extent thus the modes we are concerned with have effective radial wavenumber $k_r \lesssim k_z$. Since the critical \Af frequency rises slower than linearly with increasing $k_z$, the critical magnetic field required to shut off the instability decreases as $k_z$ increases. 

\clearpage
\clearpage

\subsection{Density Variation}
\label{denvar}
We now take for our density profile a form similar to equation (\ref{rotprof}):
\beqn{densprof}
 \rho(r) = \frac{\rho_0}{1+((r-r_{\rho})/a)^{q}},\ \ \  r>r_{\rho} \ \ ; \ \ \rho=\rho_0, \ \  \ \ r<r_{\rho} 
\eeqn
This profile, like the rotation profile, was chosen to yield constant density below $r_{\rho}$, and tend to a power law for large r. As noted above, power law density profiles have a stabilizing effect on the global modes.  The fall off of density for large radius has the effect of smoothly transitioning the effective potential to that of a locally stable vacuum magnetic field (as $\rho \to 0,\ k_2\to k_z$, and the radial solution becomes $K_1(k_z r)$). 

\begin{figure}
\plottwo{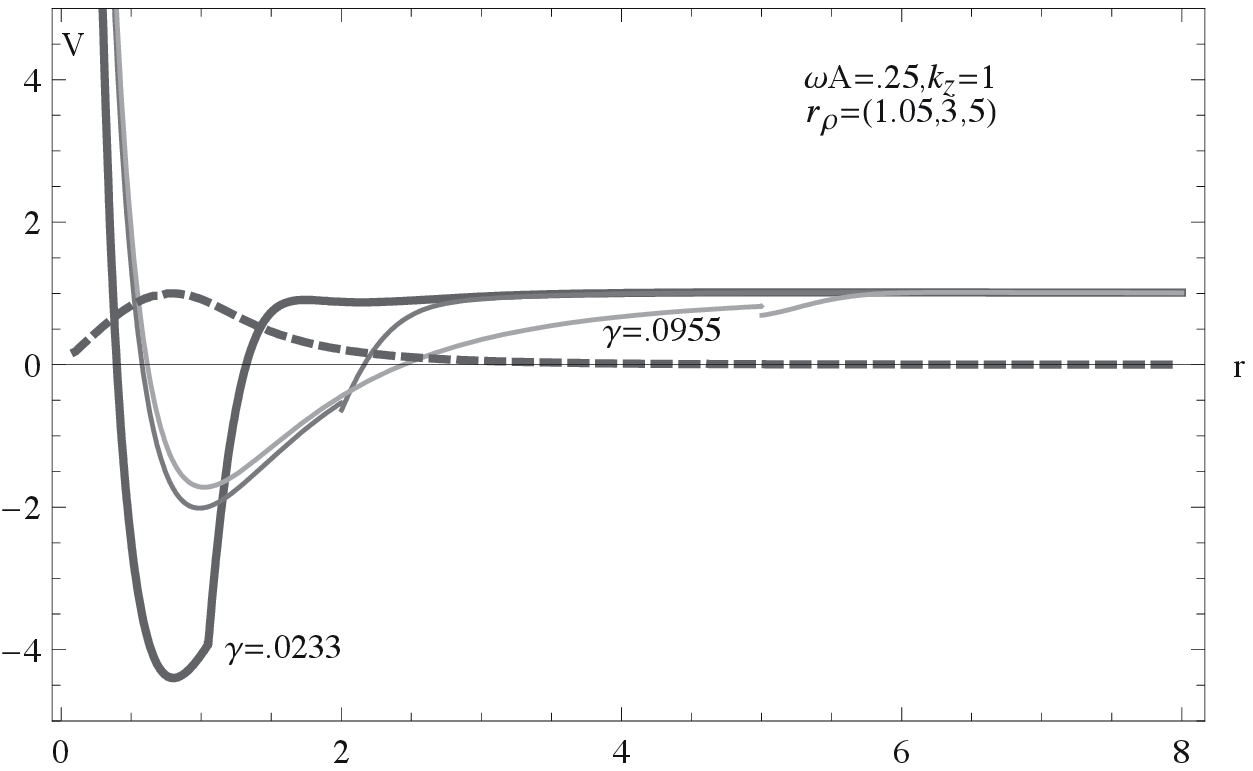}{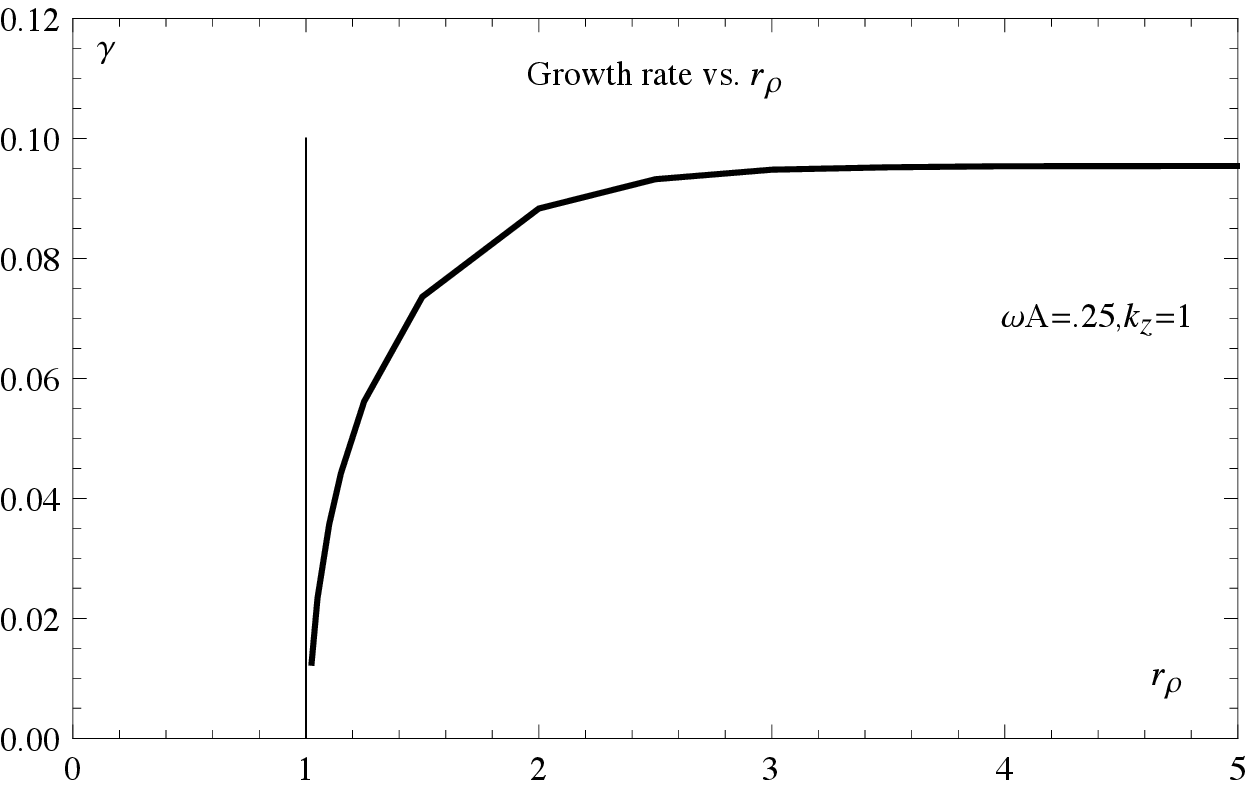}
\caption{For the profile discussed in Section \ref{denvar}, we plot the effective potential of the most unstable mode for three values of the transition radius $r_{\rho}=\{5,3,1.05\}$. 
When $r_{\rho}$ is larger than the location of maximum shear, the growth rate remains unchanged ($\gamma=.0955$ when $r_{\rho}=5$. As the density falloff moves nearer to this point, the $n=0$   growth rate decreases ($\gamma=.0233$ for $r_{\rho}=5$), and mode becomes more spatially localized (compare to Fig. \ref{effpotkep1}). The mode is eventually cut off for $r_{\rho}\simeq  1$. }
\label{denped}
\end{figure}

 To demonstrate these effects, we examine the maximum growth rates as the transition radius $r_{\rho}$ is varied, for fixed $a =R_0,\ q=2$. Figure \ref{denped} shows the effective potential for the most unstable mode as the density transition point $r_{\rho}$ is moved inward. When $r_{\rho}$ is much larger than the radial peak of the constant density mode, there is little effect on the mode, as the density is roughly constant over the region where the mode is oscillatory. As the pedestal width shrinks, the effective \Af velocity increases over the region where the mode is nonzero, raising the outer edge of the potential well. For fixed central \Af speed $\bar{v}_A$, the frequency of the mode must decrease so that the well remains deep enough to support a mode, and the peak moves inward. For $r_{\rho}$ below $\sim 1$, there is no longer a possibility of an unstable eigenmode. 

\clearpage
\clearpage

\section{Conclusion}

When linear perturbations of an inhomogeneous medium have wavelengths comparable to the equilibrium variation scale, spatial Fourier analysis becomes suspect, and global methods are more applicable. We have shown that the study of long radial wavelength incompressible axisymmetric perturbations of a differentially rotating plasma gives rise to a effective potential problem with two classes of boundary dependence. In the first class, the effective potential is negative up to the boundaries, i.e. the plasma boundaries are locally MRI-unstable. The solutions in this situation will always depend strongly on the type of boundary conditions imposed. In the case of pure Keplerian rotation, the eigenmode equation has an essential singularity at the origin.  Physically, this means that the rotational shear is maximized on the inner boundary, making the most unstable modes  ``wall'' modes, discretized by the imposed boundary. The second situation arises when the equilibrium profile is such that local MHD-stability holds at the boundaries. This can happen if the rotation shear vanishes for small radius, such as for boundary layer near the central object of an accretion disc. The spatial region over which the unstable modes exist are limited by the equilibrium flow and density profiles, leading to reduced dependence on the boundaries. The depth of the potential well is a decreasing function of growth rate.  As the local limit is achieved when the potential well is deepest, this result suggests that global modes may be more unstable.   In addition, growth rates depend on the background magnetic field in a complex fashion.  For a given perpendicular wavenumber, there exists a critical $\wa$ above which no unstable linear modes are supported, as in the case of the local MRI, but the cutoff values depend on the global properties of the density and shear flow profiles. 

We have not considered the stability of the system to global non-axisymmetric perturbations, as the effective potential treatment not as readily applicable when $m\neq0$. The potential becomes complex, and overstable convective modes can occur. These may have larger growth rates than the axisymmetric modes considered above. For similar reasons, the case of mixed toroidal and axial fields was not considered. In resistive magnetofluids, complex axisymmetric disturbances can manifest as Helical MRI modes which convect along the $\hat{z}$ axis \citep{Rud06}. Ideal MHD  Helical MRI for some simple equilibrium profiles were considered in \citet{curry95}, but these modes depend strongly on the boundary conditions. An extension to the potential theory described above will be used to study the existence and structure of these modes in a forthcoming paper.

The Authors wish to thank Dr. Richard Hazeltine and Dr. J. Craig Wheeler for useful discussions.

\end{document}